# The Origin of Corporate Control Power [*]

Jie He Min Wang [†]

**Abstract**

We discover that the probability of a shareholder possessing optimal control power evolves in Fibonacci sequence pattern and emerges as the simple harmonic oscillation $\frac{1}{2}-\frac{2}{3}-\frac{1}{2}$ in a cycle of 12 iterations. This novel feature suggests the efficiency of the allocation of corporate shareholders' right and power. Data on the Chinese stock market support this prediction.

[*] We thank the National Natural Science Foundation of China (Grant Numbers 71572149) for research support.

[†] Jie He: Professor, School of Business, Southwestern University of Finance and Economics, E-mail: hj1967way@swufe.edu.cn. Min Wang: Associate Professor, School of Business, Lin Yi University, E-mail: wangmin@lyu.edu.cn.

## 1. Introduction

A corporate shareholder has a pair of matching rights: residual claim (cash flow right) and residual right of control/control right (voting right; Alchian and Demsetz 1972; Grossman and Hart 1986; Hart and Moore 1990; Milgrom and Roberts 1992). Many previous works have contributed to the empirical correlation between this pair of rights and corporate behaviors/performance (La Porta et al. 1999; Claessens et al. 2000; Faccio and Lang 2002; Claessens et al. 2002; Lemmon and Lins 2003; Lin et al. 2013).

However, from the perspective of corporate shareholder, the exercise of voting right lies with the comparative advantage of one's voting right over the right of others rather than the absolute amount of the voting right itself. It depends not only on the amount of one's voting right itself but also that of others. It is endogenous from the competition among shareholders. That is, it is exactly "power" rather than "right". We define it as "**control power**" corresponding to "control right". No matter in logic or actuality, "control power" is the primary concept and "control right" is derivative (aroused) (Demsetz 1983), though the former is invisible, and the latter is concrete. It is the shareholder's control power, not control right, that directly dominates corporate behaviors/performance. For understanding corporate behaviors/performance, control power is more simply and more substantive than control right.

Several prior works have chosen the Shapley-Shubik power index ($SPI \in [0,1]$; Shapley and Shubik 1954) as an alternative for voting right to improve the measurement of control right (e.g., Nenova 2003; Edwards and Weichenrieder 2009; Andres et al. 2013; Aminadav and Papaioannou 2020). However, $SPI$ is an instrument to measure power, and control right should only be measured by voting right exactly.

In the endogeny of corporate shareholder's control power, it is critical to understand the dynamics. Unfortunately, a complete theoretical model that specifies how control power arises is still largely missing. Hence, we present a framework that quantifies how control power is determined and emerge explicitly.

**2. Theoretic solution and testable hypothesis**

On the fundamental principles of economics, we will derive the optimization condition of corporate shareholder's control power and then develop the testable hypothesis. The deduction is subject to identifying and understanding of the entrepreneurship of corporate shareholders (Knight 1921; Schumpeter 1912). Any one of the corporate large shareholders would be bound to compete for one's optimal control power to identify one's entrepreneurship to the greatest extent among the other shareholders.

For a corporate shareholder, we could have

$$P_r = V_r - C_r \tag{1}$$

where $r$ is the shareholder's control power, $P_r$ is the profit of $r$, $V_r$ is the revenue of $r$, and $C_r$ is the cost of $r$. With $r$ increasing, $\partial C_r/\partial r$ ($\partial V_r/\partial r$) increases (decreases) by degrees.

Moreover, we could have

$$C_r = C_{S_r} + C_{E_r} \tag{2}$$

where $S_r$ is the shares of the shareholder, $E_r$ indicates the efforts the shareholder must make to communicate with other shareholders (to affect and prevail on them so they give up the challenge/struggle for control power), $C_{S_r}$ is the opportunity cost of having shares $S_r$, and $C_{E_r}$ is the negative utility by $E_r$. Given $C_r$ unchanged, $S_r$ and $E_r$ would be changed coordinate in the opposite direction. Along with the decreasing of $S_r$, $\partial E_r/\partial S_r$ increases by degrees (see Figure 1).

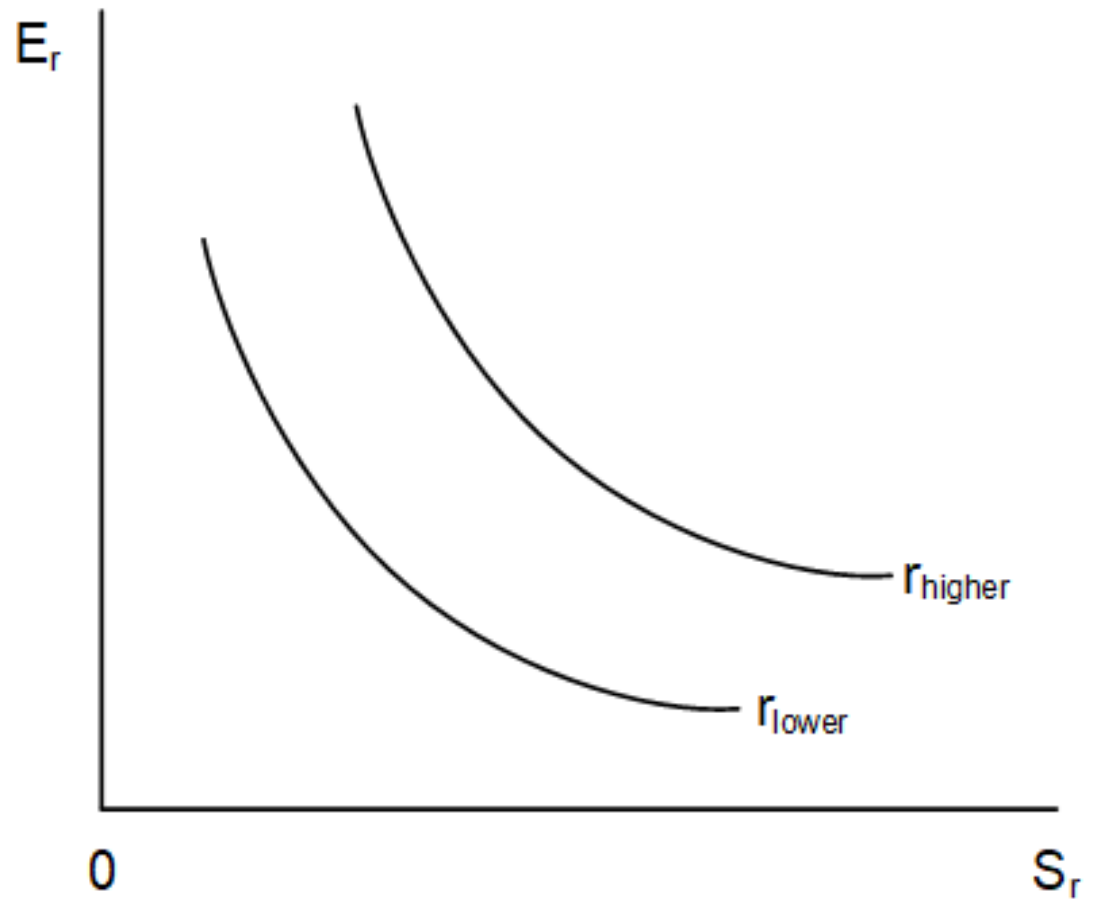

Figure 1. Coordinate changing of $S_r$ and $E_r$ in the opposite direction.

Corresponding to $E_r$, to $E_r'$- the efforts the shareholder able to make to communicate with other shareholders, along with the decreasing of $S_r$, $\partial E_r'/\partial S_r$ decreases by degrees; And along with the increasing of the entrepreneurship of the shareholder ( $e$ ), $\partial E_r'/\partial S_r$ increases by degrees.

To maximize $P_r$, the shareholder would take control power $r$ at (see Figure 2):

$$\partial V_r/\partial r = \partial C_r/\partial r \tag{3}$$

$C_r$ is then determined.

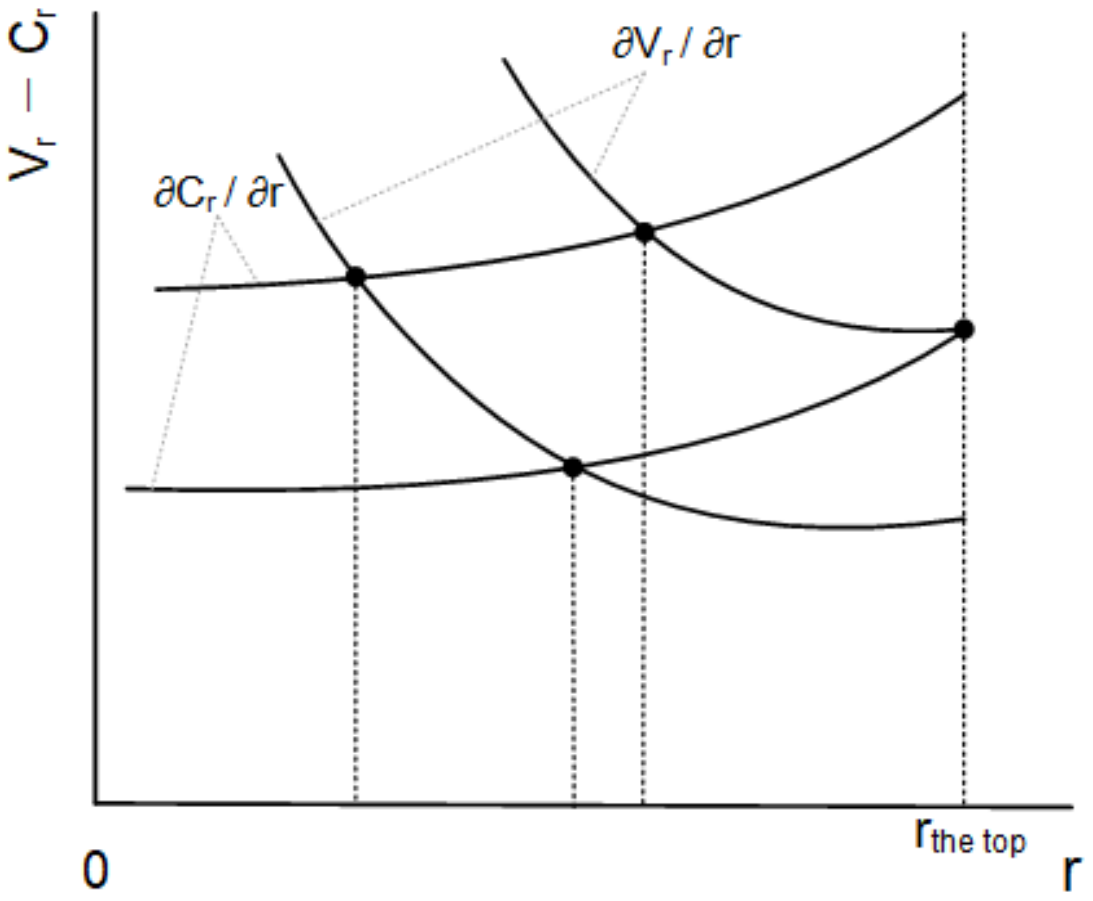

Figure 2. The shareholder maximizing his profit at $\partial V_r/\partial r = \partial C_r/\partial r$.

Of his entrepreneurship, the shareholder would be most sensitive to the positive utility of control power, and least sensitive to the negative utility of efforts to negotiate and coordinate with other shareholders. Accordingly, we predict that $r$ would reach the top value with the highest probability (see Figure 2).

Furtherly, to be identified with his entrepreneurship by the other shareholders to the greatest extent, the shareholder will take $S_r^*$ (see Figure 3), where:

$$\partial E_r/\partial S_r = \partial E_r{}'/\partial S_r \tag{4}$$

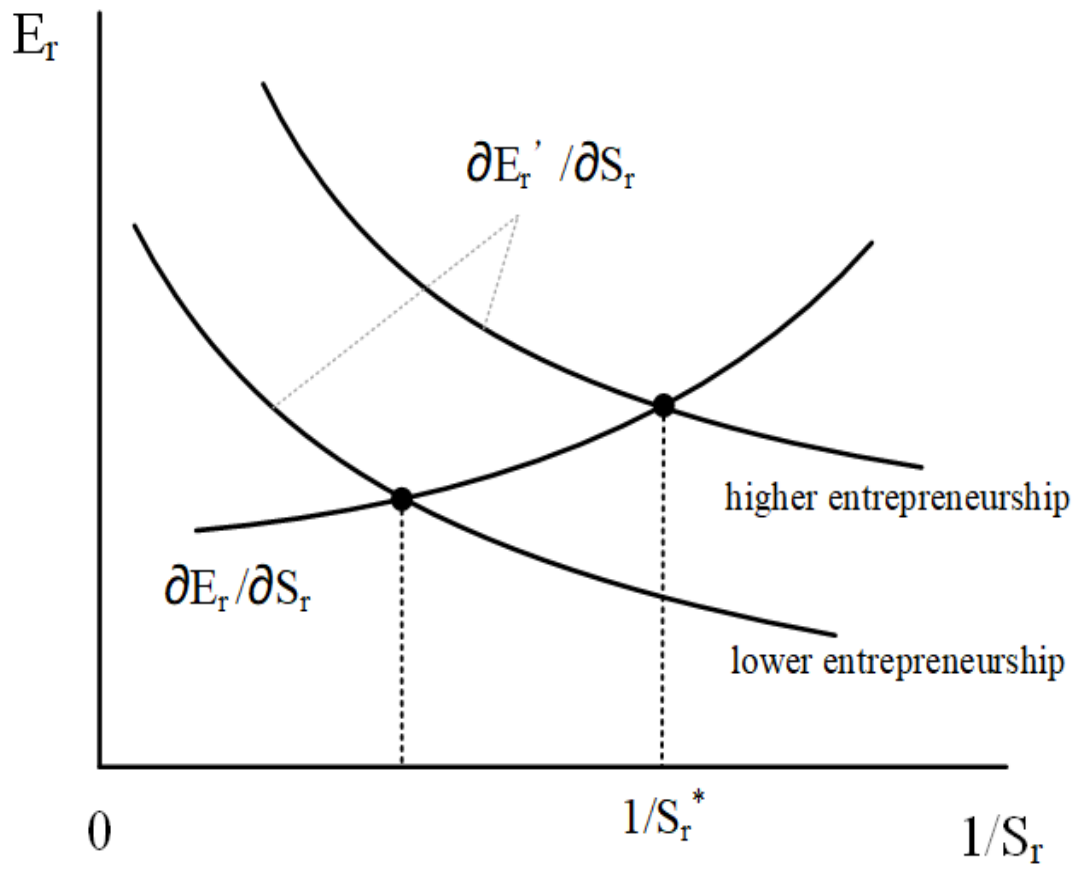

Figure 3. The shareholder is identified with his entrepreneurship by the other shareholders to the greatest extent at $\partial E_r/\partial S_r = \partial E_r'/\partial S_r$.

The economic implication of this is that, to win optimal control power, the greater the shareholder's entrepreneurship is, the few shares he could stake (mortgage) to other shareholders. For the perspective of other shareholders, it means that the weaker their entrepreneurship is, the more shares they are willing to entrust to the shareholder. It also means that the shareholder's entrepreneurship is found and used to the greatest extent and meanwhile the capital of the shareholders is used to the greatest extent too.

That is, generally, to win optimal control power, the greater the comparative advantage of entrepreneurship over that of other shareholders (over the greatest one of that) the shareholder has ($R_e^*$), the less the comparative advantage of shares over that of the other shareholders (over the sum of their shares) the shareholder would get ($R_s^*$).[3]

---

[3] Namely,

Namely, $R_e^*$ , $R_s^*$ is reciprocal in competing for optimal control power. Thus, we could have:

$$\begin{pmatrix} R_e^* \\ R_s^* \end{pmatrix}^T \begin{pmatrix} 0 & 1 \\ 1 & 0 \end{pmatrix} \begin{pmatrix} R_e^* \\ R_s^* \end{pmatrix} = \varphi \tag{5}$$

$$R_e^* * R_s^* = \varphi/2 \tag{6}$$

where φ is a power-related invariant.

We can understand this that the control power originates from the cumulative effect of $R_e^*$ on $R_s^*$.

Furtherly, for function（6）, we can deduce that there exist three sets of solution:

i. $R_e^* > 1, R_s^* > 1.$

ii. $R_e^* < 1, R_s^* \gg 1.$

iii. $R_e^* \gg 1$ , $R_s^* < 1$ . This set implicates that the shareholder with the highest entrepreneurship rather than the largest shares would be the one with maximal control power. It predicts the rise of dual-class share structure (Aggarwal et al., 2022).

---

$$\begin{pmatrix} e_0 & 0 \\ 0 & s_0 \end{pmatrix} \begin{pmatrix} R_e \\ R_s \end{pmatrix} = \begin{pmatrix} 1 & 0 \\ 0 & 1 \end{pmatrix} \begin{pmatrix} e_1 \\ s_1 \end{pmatrix}$$

Therefore,

$$\begin{pmatrix} R_e \\ R_s \end{pmatrix} = \begin{pmatrix} 1/e_0 & 0 \\ 0 & 1/s_0 \end{pmatrix} \begin{pmatrix} e_1 \\ s_1 \end{pmatrix}$$

where $e_{1,}$ $s_1$ is the entrepreneurship and the shares of a shareholder; $e_0$, $s_0$ is that of the other shareholders. We can understand that $e_1, e_0, R_e$ are intensive properties and measurement unavailable; $s_1$, $s_0$, $R_s$ are extensive properties and measurement available.

These indicate the efficiency of the allocation of right and power among corporate shareholders with different patterns in different conditions.

However, great entrepreneurship comes with great uncertainty. There is no measurement available for the entrepreneurship. Therefore, $R_e^*$ could not be calculated. Consequently, $R_s^*$ could not be calculated, and then $S_r^*$ could not be calculated.

Therefore, the shareholder could not fix $S_r^*$. What he could do is just to change $S_r$ and $E_r$ coordinate in the opposite direction ceaselessly while keeping $C_r$ unchanged. Meanwhile, the other holders will change their shares along with that. Namely, the shareholder would adjust $S_r$ and $E_r$ through updating the belief to one another's entrepreneurship by Bayes Rule ceaselessly. Logically, therefore, the shareholder' control power should be in the superimposed state of $r$/not $r$ (in the indistinguishable state of $r$/not $r$). Subject to this condition, the shareholder competes for $r$. That should be formulated as the superposition of the operations as follows:

$$\left(\begin{pmatrix}1 & 0\\0 & 0\end{pmatrix}+\begin{pmatrix}0 & 0\\1 & 0\end{pmatrix}\right)+\begin{pmatrix}0 & 1\\0 & 0\end{pmatrix}=\begin{pmatrix}1 & 1\\1 & 0\end{pmatrix} \tag{7}$$

It is the intrinsic dynamics of the evolution of the probability of a shareholder obtaining $r$. It is just the iterative matrix of Fibonacci sequence model. In Fibonacci sequence model, a pair of adult rabbits always reproduces a pair of infants, and the pair of infants always grows into a pair of adults. This describes the evolution of the adults and infants in an animal population. The dynamics is not only sufficient but also necessary for the adults to have the highest ratio in the population. The evolution function is as follows:

$$F(a)=\begin{pmatrix}1 & 1\\1 & 0\end{pmatrix}^a\begin{pmatrix}1\\1\end{pmatrix} \tag{8}$$

Where， $\begin{pmatrix}1\\1\end{pmatrix}$ is the initial state in which the probability of a shareholder obtaining $r$ is $\frac{1}{1+1}=\frac{1}{2}$.

$a = 0, 1, 2, 3, \ldots\ldots$, defined in natural number.

Equation (8) describes the evolution under the certain condition (on natural basis). To understand it subject to the uncertainty, we conduct analytic continuation to equation (8) to describe the evolution process in higher dimension which corresponds with the shareholders' practical actions. We define $a = a_\rho + a_\tau$ in real domain, $a \in (0, +\infty)$; $a_\rho \geq 0$ is defined in positive integer domain, $a_\tau \in (0,1)$ is defined in real domain.

Namely,

$$
\begin{aligned}
F(a) &= \begin{pmatrix}1 & 1\\1 & 0\end{pmatrix}^{a}\begin{pmatrix}1\\1\end{pmatrix}\\
&= \begin{pmatrix}1 & 1\\1 & 0\end{pmatrix}^{a_\rho + a_\tau}\begin{pmatrix}1\\1\end{pmatrix}\\
&= \begin{pmatrix}1 & 1\\1 & 0\end{pmatrix}^{a_\tau}\begin{pmatrix}1 & 1\\1 & 0\end{pmatrix}^{a_\rho}\begin{pmatrix}1\\1\end{pmatrix} \qquad (9)
\end{aligned}
$$

It means that the fundamental function $\begin{pmatrix}1 & 1\\1 & 0\end{pmatrix}^{a}$ would be expressed as $\begin{pmatrix}1 & 1\\1 & 0\end{pmatrix}^{a_\rho}$ by taking $\begin{pmatrix}1 & 1\\1 & 0\end{pmatrix}^{a_\tau}$ as base space. And we could understand $\begin{pmatrix}1 & 1\\1 & 0\end{pmatrix}^{a_\tau}$ as the fractal structure with quintic symmetry (quasi-periodic structure), which dimension is $a_\tau$.

Furtherly, we understand $e^{i\frac{\pi}{2}a_\tau}$ as the higher dimension structure. We would have

$$e^{i\frac{\pi}{2}a_\tau} = cos\left(\frac{\pi}{2}a_\tau\right) + i\,sin\left(\frac{\pi}{2}a_\tau\right)$$

Namely, onto observable lower dimension, $e^{i\frac{\pi}{2}a_\tau}$ is projected as $cos\left(\frac{\pi}{2}a_\tau\right)$ with the probability $p = cos^2\left(\frac{\pi}{2}a_\tau\right)$ . $p$ follows a uniform distribution $p \sim U(0,1)$ (the probabilitydensity function $f(p) = 1,\ E(p) = 1/2,\ {\sigma_p}^2 = 1/12$) to describe the uncertainty.

Furtherly, we would have:

$$p = cos^2\left(\frac{\pi}{2}a_\tau\right) \Longrightarrow \sqrt{p} = cos\left(\frac{\pi}{2}a_\tau\right) \Longrightarrow\ a_\tau = \frac{2}{\pi}cos^{-1}\sqrt{p}\ .$$

Therefore, we could redefine the fundamental evolution equation as:

$$H(p) = \begin{pmatrix} 1 & 1 \\ 1 & 0 \end{pmatrix}^{\frac{2}{\pi}cos^{-1}\sqrt{p}} \begin{pmatrix} 1 \\ 1 \end{pmatrix} \tag{10}$$

Obviously, $p \sim U(0,1)$, the description of the uncertainty, is related to the shareholder's entrepreneurship. This implicates that, in essence, exactly the finding and using of the shareholder's entrepreneurship drives the evolution.

Then, $\frac{2}{\pi}cos^{-1}\sqrt{p}$ would emerge as 0 or 1 randomly. Namely,

$$H(p) = \begin{pmatrix} 1 & 1 \\ 1 & 0 \end{pmatrix}^{\frac{2}{\pi}cos^{-1}\sqrt{p}} \begin{pmatrix} 1 \\ 1 \end{pmatrix}$$

$$= \begin{pmatrix} 1 & 1 \\ 1 & 0 \end{pmatrix}^{(0\vee1\vee0)\vee(0\vee1\vee1\vee0)\vee(0\vee1\vee1\vee1\vee0)\vee(0\vee1\vee1\vee1\vee1\vee0)\vee\cdots\cdots} \begin{pmatrix} 1 \\ 1 \end{pmatrix}$$

$$= \left\{ \begin{matrix} \left[\begin{pmatrix} 1 & 0 \\ 0 & 1 \end{pmatrix} \vee \begin{pmatrix} 1 & 1 \\ 1 & 0 \end{pmatrix}^1 \vee \begin{pmatrix} 1 & 0 \\ 0 & 1 \end{pmatrix}\right] \vee \left[\begin{pmatrix} 1 & 0 \\ 0 & 1 \end{pmatrix} \vee \begin{pmatrix} 1 & 1 \\ 1 & 0 \end{pmatrix}^2 \vee \begin{pmatrix} 1 & 0 \\ 0 & 1 \end{pmatrix}\right] \vee \\ \left[\begin{pmatrix} 1 & 0 \\ 0 & 1 \end{pmatrix} \vee \begin{pmatrix} 1 & 1 \\ 1 & 0 \end{pmatrix}^3 \vee \begin{pmatrix} 1 & 0 \\ 0 & 1 \end{pmatrix}\right] \vee \left[\begin{pmatrix} 1 & 0 \\ 0 & 1 \end{pmatrix} \vee \begin{pmatrix} 1 & 1 \\ 1 & 0 \end{pmatrix}^4 \vee \begin{pmatrix} 1 & 0 \\ 0 & 1 \end{pmatrix}\right] \vee \cdots\cdots \end{matrix} \right\} \begin{pmatrix} 1 \\ 1 \end{pmatrix}$$

$$= \left[\binom{1}{1} \vee \binom{2}{1} \vee \binom{1}{1}\right] \vee \left[\binom{1}{1} \vee \binom{3}{2} \vee \binom{1}{1}\right] \vee \left[\binom{1}{1} \vee \binom{5}{3} \vee \binom{1}{1}\right] \vee \left[\binom{1}{1} \vee \binom{8}{5} \vee \binom{1}{1}\right] \vee \cdots\cdots \tag{11}$$

Accordingly, the basic pattern of the probability evolution of corporate control power would be: $\left(\frac{1}{2} \vee \frac{2}{3} \vee \frac{1}{2}\right) \vee \left(\frac{1}{2} \vee \frac{3}{5} \vee \frac{1}{2}\right) \vee \left(\frac{1}{2} \vee \frac{5}{8} \vee \frac{1}{2}\right) \vee \left(\frac{1}{2} \vee \frac{8}{13} \vee \frac{1}{2}\right) \vee \cdots\cdots$.

Furtherly, traversing $p \sim U(0,1)$, the probability evolution equation of corporate control power would be:

$$J(p) = \frac{(2/3+1/2)}{2} + \frac{(2/3-1/2)}{2} e^{i2\pi p}$$
$$= \frac{(2/3+1/2)}{2} + \frac{(2/3-1/2)}{2} e^{i\frac{2\pi}{lcm(n)} \cdot lcm(n) \cdot p} \tag{12}$$

Here $lcm(n)$ is the least common multiple of the number (1, 2, 3, $4 \cdots \cdots n$) of iterations of $\begin{pmatrix} 1 & 1 \\ 1 & 0 \end{pmatrix}$. Let $lcm(n) \cdot p = a$ and periodically extended to $a \in (0, +\infty)$. Therefore,

$$J(a) = \frac{(2/3+1/2)}{2} + \frac{(2/3-1/2)}{2} e^{i\frac{2\pi}{lcm(n)} \cdot a}$$
$$= \frac{(2/3+1/2)}{2} + \frac{(2/3-1/2)}{2} \left(cos\left(\frac{2\pi}{lcm(n)} a\right) + i\, sin\left(\frac{2\pi}{lcm(n)} a\right)\right) \tag{13}$$

It is the probability evolution equation of a shareholder possessing optimal control power under the condition of the uncertainty.

$J(a)$ expands over time $t = h * a$. $h$ is the time interval between two operations of the evolution, which is a random variable ($> 0$). We would have:

$$G(t) = \frac{(2/3+1/2)}{2} + \frac{(2/3-1/2)}{2} \left(cos\left(\frac{2\pi}{lcm(n)} * \frac{t}{h}\right) + i\, sin\left(\frac{2\pi}{lcm(n)} * \frac{t}{h}\right)\right) \tag{14}$$

Furtherly, $8/13$ is so closely to the limit ($\approx 0.618$) that is statistically indistinguishable ((8/13)/0.618≈0.995). We could understand that the evolution sequence converges to $8/13$.

Therefore, we could have $lcm(4) = 12$ (the least common multiple of 1, 2, 3, 4). We would have:

$$G(t)_1 = \frac{(2/3+1/2)}{2} + \frac{(2/3-1/2)}{2}\left(cos\left(\frac{2\pi}{12h}t\right) + i\,sin\left(\frac{2\pi}{12h}t\right)\right) \quad (15)$$

Hence, we infer the testable hypothesis:

The ratio of the companies in which a shareholder has obtained one's optimal control power could be fitted with $G(t)_1 = \frac{(2/3+1/2)}{2} + \frac{(2/3-1/2)}{2}\left(cos\left(\frac{2\pi}{12h}t\right) + i\,sin\left(\frac{2\pi}{12h}t\right)\right)$ by Fourier function on $t$ in first order.

## 3. Approaches

### *3.1 Measurement of shareholder's control power*

The $SPI$ ($SPI \in [0,1]$) was defined as the ratio between:

i. the frequency count of a shareholder acting as a pivotal voter ("being critical to the success of a winning coalition", namely "turning a possible defeat into a success"; Shapley and Shubik 1954), and

ii. the total number of possible coalitional combinations among the shareholders vying for the power.

Economically, this indicates the probability of a shareholder being the last decision-maker in all competitor coalitions. That is exactly what we want to describe for shareholders' control power. We use $SPI$ to measure it.

*3.1.1 Calculation of the value of $SPI$.* We use the program *ssdirect*[4] to calculate the value of $SPI$ of the shareholders.

*3.1.2 Voting rule.* In the Chinese stock market, the "one share one vote" rule is applicable. In shareholders' meeting (the highest authority), by the half-win ($1/2$) principle, the board directors, members of the top decision-making body (top power body), are elected. Therefore, we take it as the voting rule to compute the shareholders' $SPI$.

*3.1.3 Which shareholders are scrambling for control power?* In a company, shareholders' meeting is the supreme authority institution. The shareholders decide fundamental issues through voting in it. As shown in Table 1, for the companies on main board in the Chinese stock market, from 2000 to 2024, the mean of the ratio of the share rate being present in shareholders' meeting to the share rate of top10 shareholders is rather close to 1.0. Further, the ratio of companies with top1 shareholder's $SPI = 1$ playing in top9, top10, top11 shareholders is rather close. Therefore, we would consider in statistics that, in Chinese stock market, shareholders' meeting is conducted almost within the top10 shareholders.

Table 1. Which shareholders are scrambling for control power?

| Year | $\frac{S_{meeting}}{S_{top10}}$ Mean | $\frac{S_{meeting}}{S_{top10}}$ S.D. | $\frac{N_{[mean-1s.d,mean+1s.d]}}{N_{sample}}$ | $R_{SPI=1(top9)}$ | $R_{SPI=1(top10)}$ | $R_{SPI=1(top11)}$ | $N_{sample}$ |
|---|---|---|---|---|---|---|---|
| 2000 | 0.946 | 0.130 | 0.890 | 0.782 | 0.775 | 0.767 | 1030 |

[4] Written by Dennis Leech (University of Warwick, UK) & Robert Leech (Imperial College London, UK). Thanks for their supports.

| | | | | | | | |
|---|---|---|---|---|---|---|---|
| 2001 | 0.944 | 0.108 | 0.865 | 0.783 | 0.774 | 0.769 | 1139 |
| 2002 | 0.931 | 0.109 | 0.846 | 0.771 | 0.765 | 0.763 | 1210 |
| 2003 | 0.933 | 0.122 | 0.851 | 0.743 | 0.739 | 0.732 | 1266 |
| 2004 | 0.923 | 0.113 | 0.836 | 0.734 | 0.726 | 0.724 | 1326 |
| 2005 | 0.889 | 0.145 | 0.820 | 0.685 | 0.676 | 0.672 | 1142 |
| 2006 | 0.890 | 0.133 | 0.775 | 0.697 | 0.688 | 0.681 | 1255 |
| 2007 | 0.886 | 0.134 | 0.805 | 0.737 | 0.727 | 0.721 | 1295 |
| 2008 | 0.876 | 0.135 | 0.804 | 0.762 | 0.754 | 0.747 | 1279 |
| 2009 | 0.871 | 0.136 | 0.807 | 0.773 | 0.764 | 0.762 | 1280 |
| 2010 | 0.870 | 0.140 | 0.808 | 0.774 | 0.769 | 0.765 | 1299 |
| 2011 | 0.869 | 0.141 | 0.826 | 0.779 | 0.770 | 0.767 | 1354 |
| 2012 | 0.872 | 0.132 | 0.800 | 0.773 | 0.766 | 0.763 | 1357 |
| 2013 | 0.870 | 0.136 | 0.806 | 0.765 | 0.758 | 0.755 | 1363 |
| 2014 | 0.877 | 0.143 | 0.815 | 0.748 | 0.740 | 0.737 | 1383 |
| 2015 | 0.853 | 0.142 | 0.801 | 0.718 | 0.708 | 0.705 | 1438 |
| 2016 | 0.851 | 0.143 | 0.781 | 0.684 | 0.670 | 0.667 | 1575 |
| 2017 | 0.851 | 0.148 | 0.785 | 0.650 | 0.637 | 0.634 | 1745 |
| 2018 | 0.845 | 0.141 | 0.770 | 0.640 | 0.626 | 0.625 | 1790 |
| 2019 | 0.850 | 0.151 | 0.806 | 0.634 | 0.627 | 0.624 | 1826 |
| 2020 | 0.851 | 0.141 | 0.768 | 0.617 | 0.612 | 0.608 | 1911 |
| 2021 | 0.857 | 0.140 | 0.765 | 0.628 | 0.621 | 0.616 | 2032 |
| 2022 | 0.862 | 0.134 | 0.762 | 0.644 | 0.634 | 0.630 | 2089 |
| 2023 | 0.865 | 0.132 | 0.769 | 0.648 | 0.634 | 0.632 | 2123 |
| 2024 | 0.879 | 0.132 | 0.793 | 0.657 | 0.648 | 0.645 | 2137 |

*Notes*:

$$S_{meeting} = \frac{share\ rate\ present\ in\ the\ annual\ meeting + share\ rate\ present\ in\ the\ extraordinary\ meeting}{number\ of\ annual\ and\ extraordinary\ shareholders'\ meetings\ in\ the\ year};$$

$S_{top10}$ = share rate of top10 shareholders at the end of year; $N_{[mean-1s.d,mean+1s.d]}$ = number of companies of $S_{meeting}/S_{top10}$ in $[mean - 1s.d, mean + 1s.d]$; $R_{SPI=1(top9)}$, $R_{SPI=1(top10)}$, $R_{SPI=1(top11)}$ = the ratio of companies with top1 shareholder's $SPI = 1$ playing in top9, top10, top11(taking $S_{meeting} - S_{top10}$ (if $< 0$, taking as 0) as the share rate of the 11th player) shareholders respectively; $N_{sample}$ = total number of companies.

*3.2 Sampling*

When the top1 shareholder's share rate is ⩾50%, his $SPI$ is 1, and no effort is needed to negotiate and coordinate with other shareholders. Therefore, our sample is composed of companies in which the top1 shareholder's share rate is < 50%.

All data for this study are derived from the database of the China Stock Market and Accounting Research.

*3.3 Grouping of the natural experiments*

In Chinese stock market, the main board is for the established companies, as well as the small and medium-sized enterprises board (SME) and growth enterprises market board (GEM) are for the start-ups. In these companies, the top1 shareholder is either privately owned or state-owned. The state-owned shareholders' decisions on adding/reducing shares often be interfered and restricted by the government[5]. However, it is difficult for the government to interfere privately

[5] Order of China Securities Regulatory Commission (CSRC) No. 86: Administrative Measures for the Reform of Non-tradable Shares in Listed Companies, September 4, 2005.

http://www.csrc.gov.cn/pub/newsite/flb/flfg/bmgf/ssgs/gqfz/201012/t20101231_189887.html

owned shareholders' actions. In SME and GME, because of lower listing requirements[6] and greater uncertainty of corporate future expected returns than in main board, the start-ups' shareholders are more likely to misconduct in corporate finance. Therefore, from the normal viewpoint of the shareholding, the share adding/reducing action of the holders of the start-ups would deviate from the normal pattern predicted by our theory (Liu and Zhao 2011; Jia 2017; He

---

Order of State-owned Assets Supervision and Administration Commission of the State Council (SASAC) and China Securities Regulatory Commission (CSRC) No. 19: Interim Administrative Measures for the State-owned Shares Transferring of Listed Companies, July 1, 2007.
http://www.csrc.gov.cn/pub/newsite/flb/flfg/bmgz/ssl/201012/t20101231_189730.html
Order of State-owned Assets Supervision and Administration Commission of the State Council (SASAC), Ministry of Finance of the People's Republic of China (MOF) and China Securities Regulatory Commission (CSRC) No. 36: Administrative Measures for the Supervision over the State-owned Equity of Listed Companies, July 1,2018.
http://www.csrc.gov.cn/pub/newsite/zjhxwfb/xwdd/201805/t20180518_338364.html

[6] Order of China Securities Regulatory Commission (CSRC) No. 32: Administrative Measures for Initial Public Offering and Listing, May 18, 2006.
http://www.csrc.gov.cn/pub/zjhpublic/zjh/200804/t20080418_14502.html
Order of China Securities Regulatory Commission (CSRC) No. 61: Interim Administrative Measures for Initial Public Offering and Listing on GEM Board, May 1, 2009.
http://www.csrc.gov.cn/pub/zjhpublic/G00306202/cyb/200911/t20091117_170416.html

2019). Thus, we take privately owned companies on the main board as the control group and focus on them by considering them as the normal firms whose shareholders' action is not distorted; and take state-owned companies on the main board as well as state-owned and privately owned companies on the SME and GEM board as the natural experiment groups (whose shareholders' action is distorted) in contrast.

*3.4 The optimal control power of corporate shareholder*

We can understand the choice of top1 shareholder's optimal control power by the distribution of $SPI$ value.

We find that the distribution of top1 shareholder's $SPI$ is not normal but rather nearly power-law. In about 46–66% of the privately owned companies on the main board, the top1 shareholder's $SPI = 1$ (Table 2). This implies that top1 shareholder's own choice is more likely to be $SPI = 1$.

Table 2. Top1 shareholder's $SPI = 1$.

| Year | Main board | | | | SME and GEM boards | | | |
|---|---|---|---|---|---|---|---|---|
| | Privately owned | | State-owned | | Privately owned | | State-owned | |
| | $N_{SPI=1}/N_{sample}$ | $N_{sample}$ | $N_{SPI=1}/N_{sample}$ | $N_{sample}$ | $N_{SPI=1}/N_{sample}$ | $N_{sample}$ | $N_{SPI=1}/N_{sample}$ | $N_{sample}$ |
| 1992 | 1.000 | 1 | 0.444 | 9 | | | | |
| 1993 | 0.500 | 28 | 0.600 | 70 | | | | |
| 1994 | 0.531 | 32 | 0.628 | 94 | | | | |
| 1995 | 0.575 | 40 | 0.615 | 117 | | | | |
| 1996 | 0.644 | 87 | 0.660 | 232 | | | | |
| 1997 | 0.561 | 114 | 0.678 | 314 | | | | |
| 1998 | 0.508 | 124 | 0.689 | 357 | | | | |
| 1999 | 0.493 | 146 | 0.668 | 391 | | | | |
| 2000 | 0.500 | 166 | 0.664 | 458 | | | | |
| 2001 | 0.489 | 184 | 0.669 | 502 | | | | |
| 2002 | 0.482 | 226 | 0.673 | 514 | | | | |
| 2003 | 0.442 | 294 | 0.675 | 508 | | | | |
| 2004 | 0.470 | 334 | 0.651 | 525 | 0.429 | 28 | 0.000 | 5 |
| 2005 | 0.463 | 337 | 0.650 | 540 | 0.459 | 37 | 0.400 | 10 |
| 2006 | 0.491 | 389 | 0.696 | 644 | 0.426 | 68 | 0.500 | 22 |
| 2007 | 0.569 | 404 | 0.725 | 661 | 0.409 | 127 | 0.590 | 39 |
| 2008 | 0.610 | 395 | 0.749 | 658 | 0.465 | 172 | 0.568 | 44 |
| 2009 | 0.664 | 402 | 0.742 | 648 | 0.483 | 238 | 0.623 | 53 |
| 2010 | 0.663 | 412 | 0.755 | 661 | 0.442 | 475 | 0.507 | 73 |
| 2011 | 0.662 | 444 | 0.750 | 651 | 0.439 | 670 | 0.565 | 85 |
| 2012 | 0.639 | 452 | 0.754 | 638 | 0.457 | 775 | 0.585 | 94 |
| 2013 | 0.630 | 465 | 0.746 | 625 | 0.506 | 785 | 0.594 | 96 |
| 2014 | 0.601 | 504 | 0.748 | 636 | 0.540 | 881 | 0.653 | 98 |
| 2015 | 0.551 | 584 | 0.731 | 651 | 0.502 | 1,016 | 0.642 | 106 |
| 2016 | 0.515 | 660 | 0.694 | 667 | 0.468 | 1,138 | 0.593 | 118 |
| 2017 | 0.475 | 827 | 0.695 | 681 | 0.446 | 1,347 | 0.548 | 126 |
| 2018 | 0.479 | 873 | 0.673 | 675 | 0.450 | 1,363 | 0.517 | 143 |
| 2019 | 0.485 | 897 | 0.669 | 683 | 0.464 | 1,404 | 0.465 | 185 |
| 2020 | 0.454 | 963 | 0.673 | 703 | 0.468 | 1,484 | 0.502 | 227 |
| 2021 | 0.466 | 1029 | 0.682 | 730 | 0.466 | 1,605 | 0.539 | 269 |
| 2022 | 0.504 | 1038 | 0.665 | 772 | 0.487 | 1,701 | 0.554 | 296 |
| 2023 | 0.506 | 1045 | 0.654 | 788 | 0.503 | 1,761 | 0.580 | 305 |
| 2024 | 0.498 | 1068 | 0.696 | 762 | 0.519 | 1782 | 0.592 | 311 |

*Notes*: $N_{SPI=1}$-number of companies with the top1 shareholder's $SPI = 1$.

Further, we find normal distribution ( $N(0.465, 0.164^2)$ ) in the sample of privately owned firms on the main board with the top1 shareholder's control power $SPI < 1$ (Table 3). This normal distribution is robust and independent of the ratio of companies with the top1 shareholder's control power $SPI = 1$. It suggests that this result has not occurred because of the top1 shareholder's own choice but rather because of a defeat in the competition for $SPI = 1$, which they can't but accept.

Table 3. Description of the sample of privately owned firms on the main board with top1 shareholder having $SPI < 1$.

| Year | Max | Min | Mean | S.D. | $\frac{N_{[mean-1s.d,mean+1s.d]}}{N_{sample}}$ | $N_{sample}$ |
|---|---|---|---|---|---|---|
| 1993 | 0.711 | 0.170 | 0.389 | 0.185 | 0.643 | 14 |
| 1994 | 0.800 | 0.221 | 0.411 | 0.173 | 0.667 | 15 |
| 1995 | 0.800 | 0.197 | 0.434 | 0.188 | 0.706 | 17 |
| 1996 | 0.800 | 0.218 | 0.450 | 0.162 | 0.710 | 31 |
| 1997 | 0.800 | 0.225 | 0.471 | 0.157 | 0.720 | 50 |
| 1998 | 0.800 | 0.224 | 0.462 | 0.157 | 0.672 | 61 |
| 1999 | 0.800 | 0.221 | 0.474 | 0.163 | 0.676 | 74 |
| 2000 | 0.800 | 0.144 | 0.474 | 0.161 | 0.687 | 83 |
| 2001 | 0.800 | 0.144 | 0.480 | 0.162 | 0.734 | 94 |
| 2002 | 0.800 | 0.147 | 0.470 | 0.157 | 0.675 | 117 |
| 2003 | 0.800 | 0.149 | 0.474 | 0.158 | 0.707 | 164 |
| 2004 | 0.800 | 0.148 | 0.462 | 0.151 | 0.712 | 177 |
| 2005 | 0.800 | 0.143 | 0.468 | 0.153 | 0.713 | 181 |
| 2006 | 0.800 | 0.143 | 0.500 | 0.165 | 0.591 | 198 |
| 2007 | 0.800 | 0.154 | 0.470 | 0.161 | 0.718 | 174 |
| 2008 | 0.800 | 0.159 | 0.470 | 0.157 | 0.701 | 154 |
| 2009 | 0.800 | 0.160 | 0.507 | 0.173 | 0.637 | 135 |
| 2010 | 0.800 | 0.169 | 0.484 | 0.170 | 0.662 | 139 |
| 2011 | 0.800 | 0.164 | 0.483 | 0.166 | 0.647 | 150 |
| 2012 | 0.800 | 0.197 | 0.478 | 0.161 | 0.669 | 163 |
| 2013 | 0.800 | 0.137 | 0.468 | 0.167 | 0.657 | 172 |
| 2014 | 0.800 | 0.137 | 0.472 | 0.177 | 0.692 | 201 |
| 2015 | 0.800 | 0.135 | 0.474 | 0.173 | 0.687 | 262 |
| 2016 | 0.800 | 0.135 | 0.464 | 0.166 | 0.675 | 320 |
| 2017 | 0.800 | 0.135 | 0.474 | 0.176 | 0.661 | 434 |
| 2018 | 0.800 | 0.135 | 0.473 | 0.172 | 0.664 | 455 |
| 2019 | 0.800 | 0.119 | 0.461 | 0.161 | 0.680 | 462 |
| 2020 | 0.800 | 0.160 | 0.459 | 0.156 | 0.669 | 526 |
| 2021 | 0.800 | 0.142 | 0.459 | 0.159 | 0.711 | 550 |
| 2022 | 0.800 | 0.160 | 0.455 | 0.155 | 0.705 | 515 |
| 2023 | 0.800 | 0.146 | 0.460 | 0.160 | 0.694 | 516 |
| 2024 | 0.800 | 0.146 | 0.453 | 0.153 | 0.698 | 536 |
| Mean<br>(S.D) | 0.797<br>(0.016) | 0.162<br>(0.031) | 0.465<br>(0.022) | 0.164<br>(0.009) | 0.683<br>(0.030) | 32 |

*Notes*: $N_{[mean-1s.d,mean+1s.d]}$ = number of companies with top1 shareholder's $SPI$ in $[mean - 1s.d, mean + 1s.d]$.

All of these indicate that, for almost every company in the Chinese stock market, the choice of the top1 shareholder's optimal control power is $SPI = 1$. This result supports our theoretical prediction above. Moreover, this gives us the opportunity to test our hypotheses using the ratio of companies with top1 shareholder that have obtained control power with $SPI = 1$, $R_{SPI=1}$.

*3.5 The time interval between the state and the state of the evolution, h*

$h$ is a random variable. When we use the data of end of the year, it would only be a positive integer as 1, 2, $\dots n$ years. Further, due to three-year election cycle of directors in listed company[7], $h$ would be more likely to be 1 or 2 years and more likely to have the probability of 1/2. Therefore, we could consider $h = 1 * \frac{1}{2} + 2 * \frac{1}{2} = 1.5$ years statistically.

## 4. Results

Using data of the Chinese stock market for 1996–2024 (sample size ≥ 50), we obtained a good fit with $R_{SPI=1}$ on $t$ with Fourier function in first order (Figure 4a). The result could not refute our hypothesis statistically.

---

[7] The Company Law of the People's Republic of China (Revision 2018). The Sixth Meeting of the Standing Committee of the 13th National People's Congress, October 26, 2018. http://www.npc.gov.cn/zgrdw/npc/xinwen/2018-11/05/content_2065671.htm.

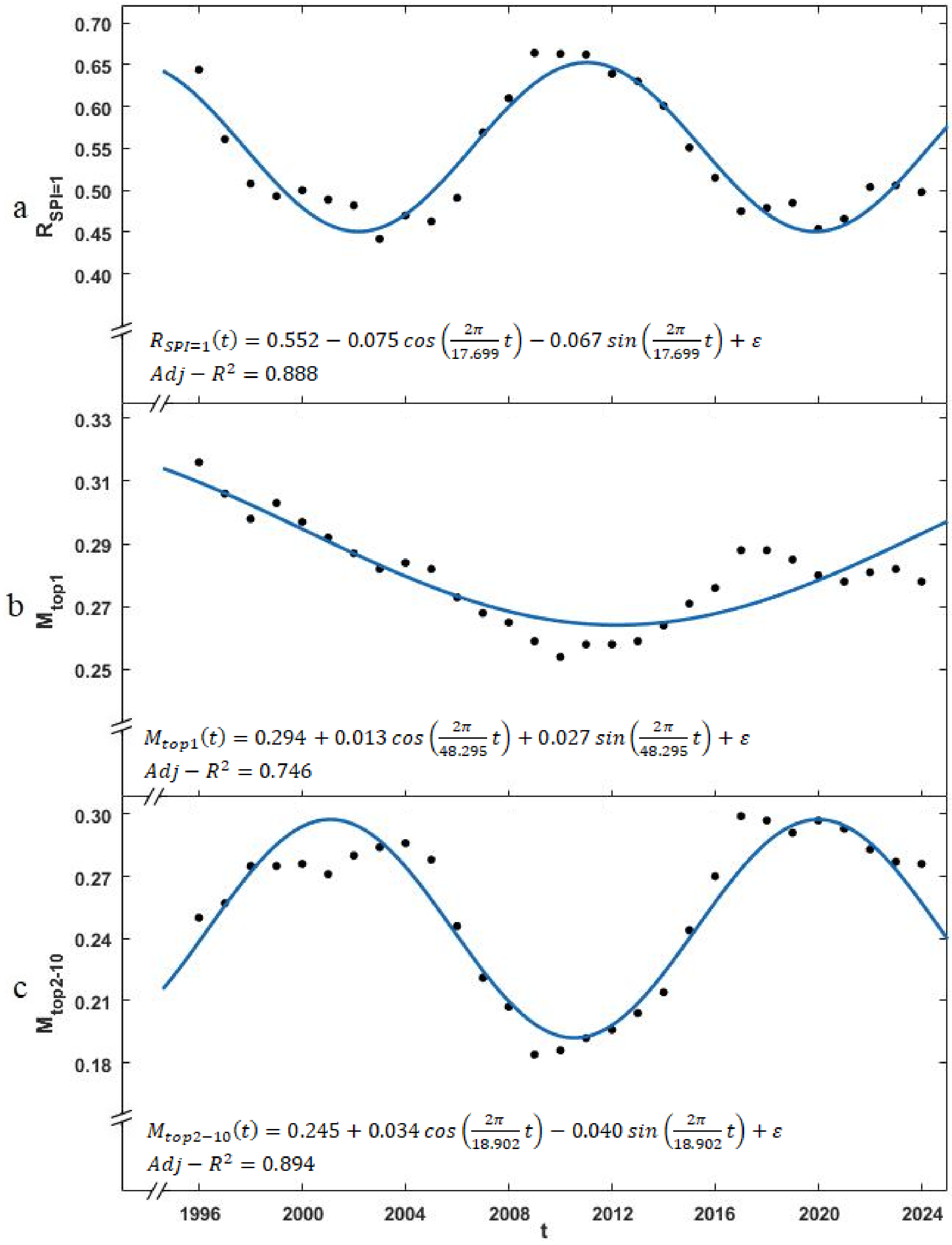

Figure 4. Fitting of $R_{SPI=1}$, $M_{top1}$, and $M_{top2-10}$ on $t$ with Fourier function in first order for privately owned firms on the main board in 1996–2024.

As shown in Figure 1, $S_r$ and $E_r$ are changed coordinate in the opposite direction. That would make the changing of the top2-10 shareholders' shares ($S_{2-10}$).

Subject to

$$S_r = A\cos\left(\frac{2\pi}{T}t + \pi\right) \tag{16}$$

$$E_r = B\cos\left(\frac{2\pi}{T}t\right) \tag{17}$$

where $t$ is time.

Then

$$\begin{aligned} S_{2-10} &\propto A\cos\left(\frac{2\pi}{T}t + \pi\right) * B\cos\left(\frac{2\pi}{T}t\right) \\ &= \frac{A*B}{2}\left[\cos\left(\frac{4\pi}{T}t + \pi\right) + \cos\pi\right] \\ &= \frac{A*B}{2}\left[\cos\left(\frac{2\pi}{T/2}t + \pi\right) - 1\right] \end{aligned} \tag{18}$$

Therefore, for the sample, we can furtherly predict:

i. The period of mean of the top2-10 shareholders' share rate ($M_{top2-10}$) on $t$ is $1/2$ of that of the top1 shareholder's share rate ($M_{top1}$).

And the comparative advantage of the top1 shareholder's voting right over the top2-10 shareholders determines that his control power ($SPI$) is equal to 1 or not. Therefore,

ii. The period of mean of the top2-10 shareholders' share rate ($M_{top2-10}$) on $t$ is equal to that of $R_{SPI=1}$ but with phase position difference $\pi$.

The results meet our predictions (Figure 4b, 4c and Appendix Table A.1).

Statistically, as the main founder of the firm, the top1 shareholder's entrepreneurship is significantly greater than that of the others. Therefore, in the subsample of the companies with the top1 holder's $SPI = 1$, we could predict that:

a. The ratio of the top1 holder's shares to the top2-10 holder's shares ($\frac{top1\ holder's\ shares}{top2-10\ holder's\ shares}$) would present power-law distribution.

And in the subsample of the companies with the top1 holder's $SPI < 1$, because of the striving and chasing for shares among the shareholders, we could predict that:

b. $\frac{top1\ holder's\ shares}{top2-10\ holder's\ shares}$ would be uniformly distributed.

The results meet our predictions (Appendix Table A.2 and Table A.3).

Only in our theoretic framework, we could deduce that the top1 shareholder's shares and the top2-10 shareholders' shares would appear as these patterns. This proves that our theory is not only sufficient but also necessary for our hypothesis.

In contrast, the fittings in the state-owned firms on the main board as well as the state-owned and privately owned firms on the SME and GEM board deviate from the patterns in the privately owned firms in the main board. The results are consistent with our predictions (Appendix Figure A.1, Table A.4, Figure A.2, A.3, Table A.5, A.6).

$R_{SPI=1}$ is not correlative to the stock market index and the economic growth of China. Moreover, $M_{top1}$ and $M_{top2-10}$ have negative correlation with the stock market index (Appendix Figure A.4 and Table A.7). This explains the independent and universal feature of $R_{SPI=1}$ and the dependence of $M_{top1}$ and $M_{top2-10}$ on the opportunity cost of shareholding.

## 5. Conclusions

Our framework for corporate shareholder's control power differs from the empirical models that fit curves to phenomenological data. In contrast, we theoretically and quantitatively deduce that the probability of a shareholder obtaining optimal control power evolves in Fibonacci sequence pattern. It is a story of right transfer and power origin among corporate shareholders. It captures the salient, dynamic, and universal features of normal firm. This specific pattern, as the benchmark, demonstrates the elemental dynamics of the shareholder's control right and power. It indicates that the shareholder's control power comes from the cumulative effect of the comparative advantage of entrepreneurship over other shareholders on the comparative advantage of shares over other shareholders. It could be understood that the nature of firm is the finding and using of entrepreneurship. It implies the efficiency of the allocation of right and power among corporate shareholders. This deduction can be applied to understand the efficiency equivalence of different patterns of allocation of right and power, such as "one share one vote", "dual-class share structure", "limited partnership", and different capital structure, and so on, under zero transaction cost condition.

Furtherly, our framework could also be extended to understand the power allocation in classical enterprise (the origin of shareholder's residual claim and residual right of control). We could unify it as follows:

Table 4. The unification of our framework to classical enterprise

| who holds power? | non-comparative advantage factor | | |
|---|---|---|---|
| | I | II | III |

| | | holding shares | entrepreneurship | labor |
|---|---|---|---|---|
| comparative advantage factor | holding shares | | yes | yes or no |
| | entrepreneurship | yes | | yes or no |
| | labor | no | no | |

Here, in a company, there are two partners each having three factors: capital (shares), entrepreneurship and labor which have different property characteristics (Table 5).

Table 5. The factors with different property characteristics

| | holding shares | entrepreneurship | labor | *unknown factor?* |
|---|---|---|---|---|
| measurement available | yes | no | yes | no |
| mortgageable | yes | no | no | yes |

Table 4 shows the control power allocation in different status of the comparative advantages and the property characteristics of the factors. The column I and II describe the power allocation in classical enterprise: capital employs labor (the shareholder has the right of residual claim and residual control-the control power over labor). The column III describes the control power allocation among shareholders studied in our paper.

Furthermore, we may generalize this theory to analyze the behaviors of politic parties.

We quantitatively confirm this in line with the data for the Chinese stock market. Nevertheless, how well it captures the fundamental feature of corporate control power in other markets such as in the US or Europe remains to be tested.

# Appendix

## Appendix Figures:

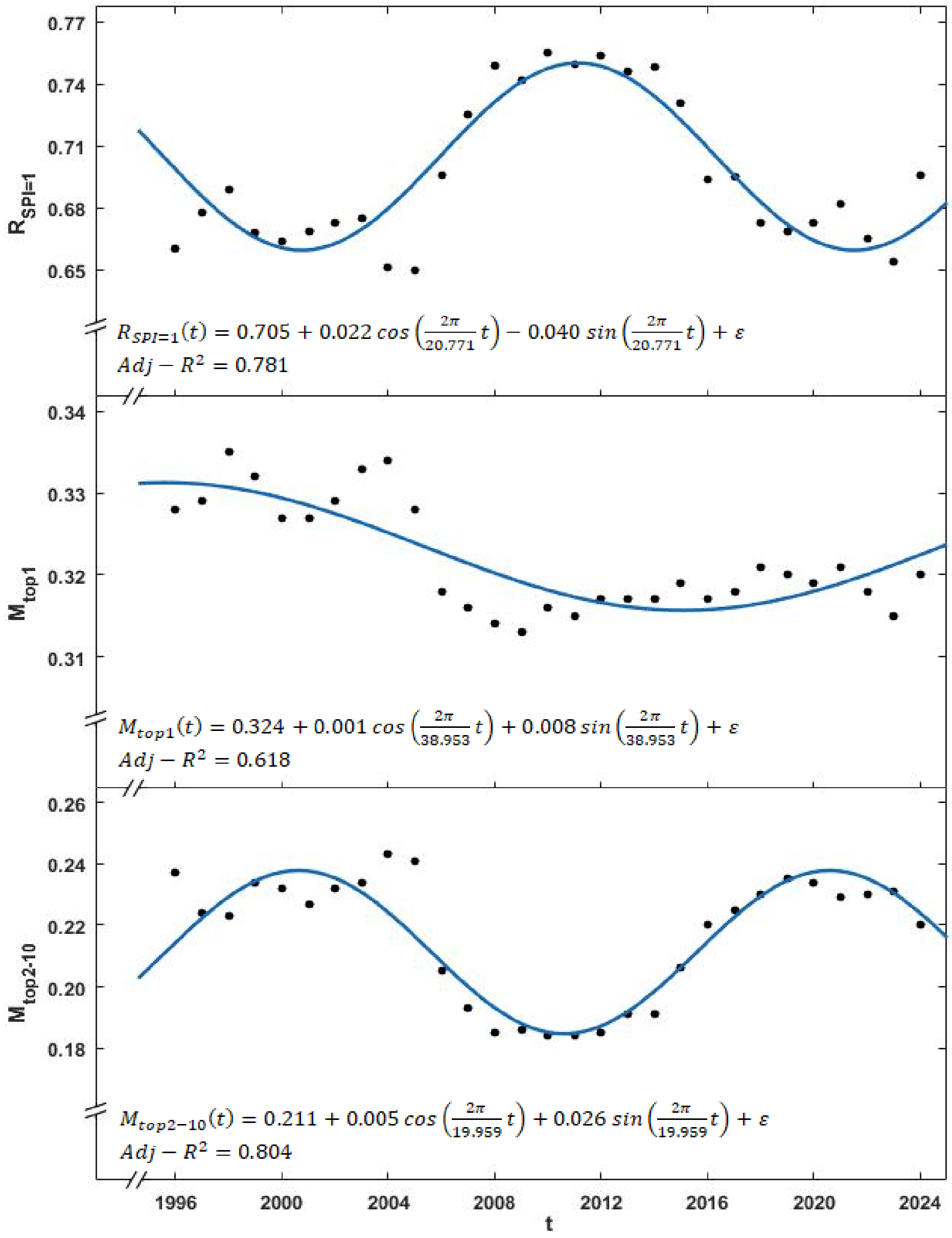

Figure A.1: Fitting of $R_{SPI=1}$, $M_{top1}$, and $M_{top2-10}$ on $t$ with a Fourier function in the first order for state-owned firms on the main board in 1996–2024.

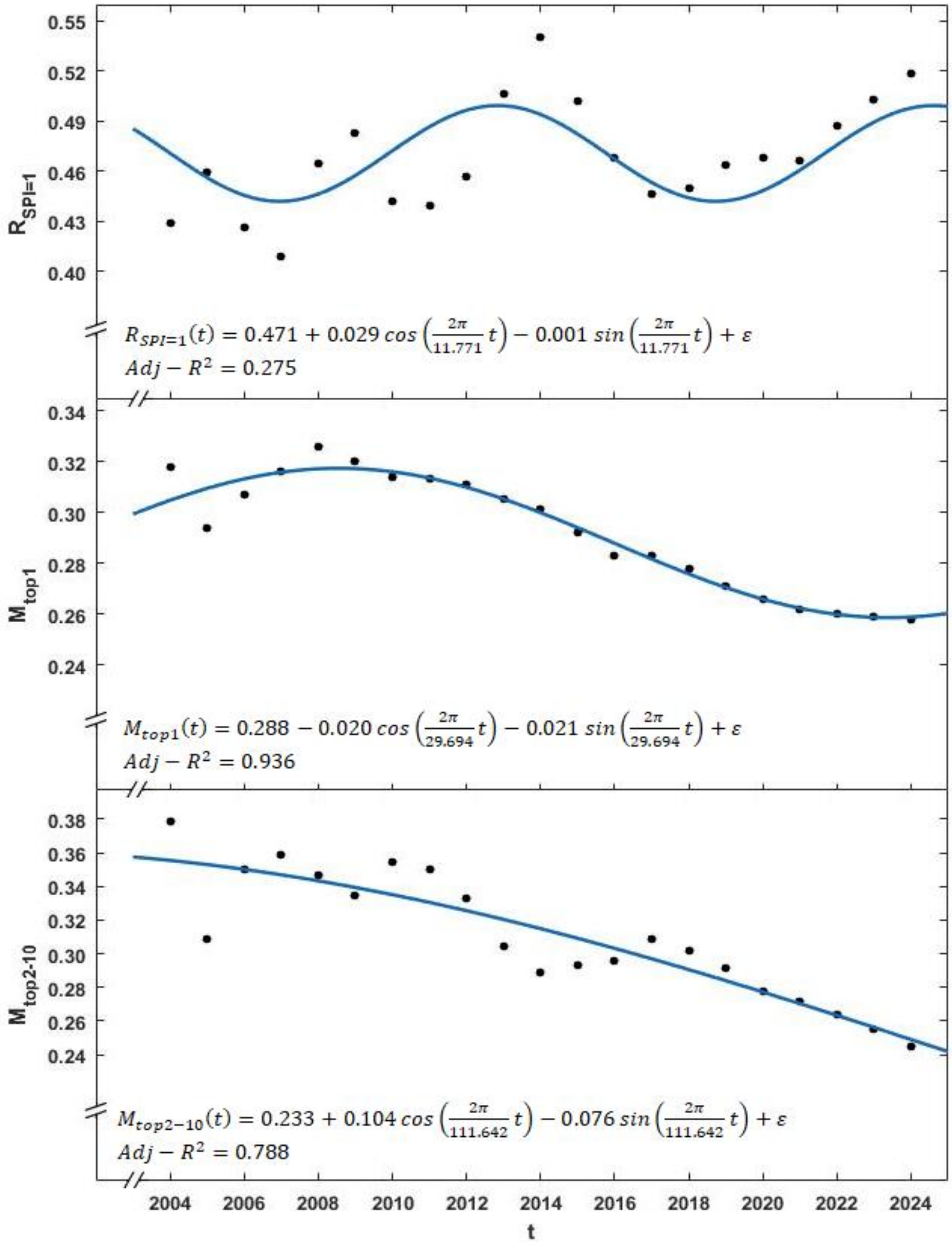

Figure A.2: Fitting of $R_{SPI=1}$, $M_{top1}$, and $M_{top2-10}$ on $t$ with a Fourier function in the first order for privately owned firms on the SME and GEM boards in 2004–2024.

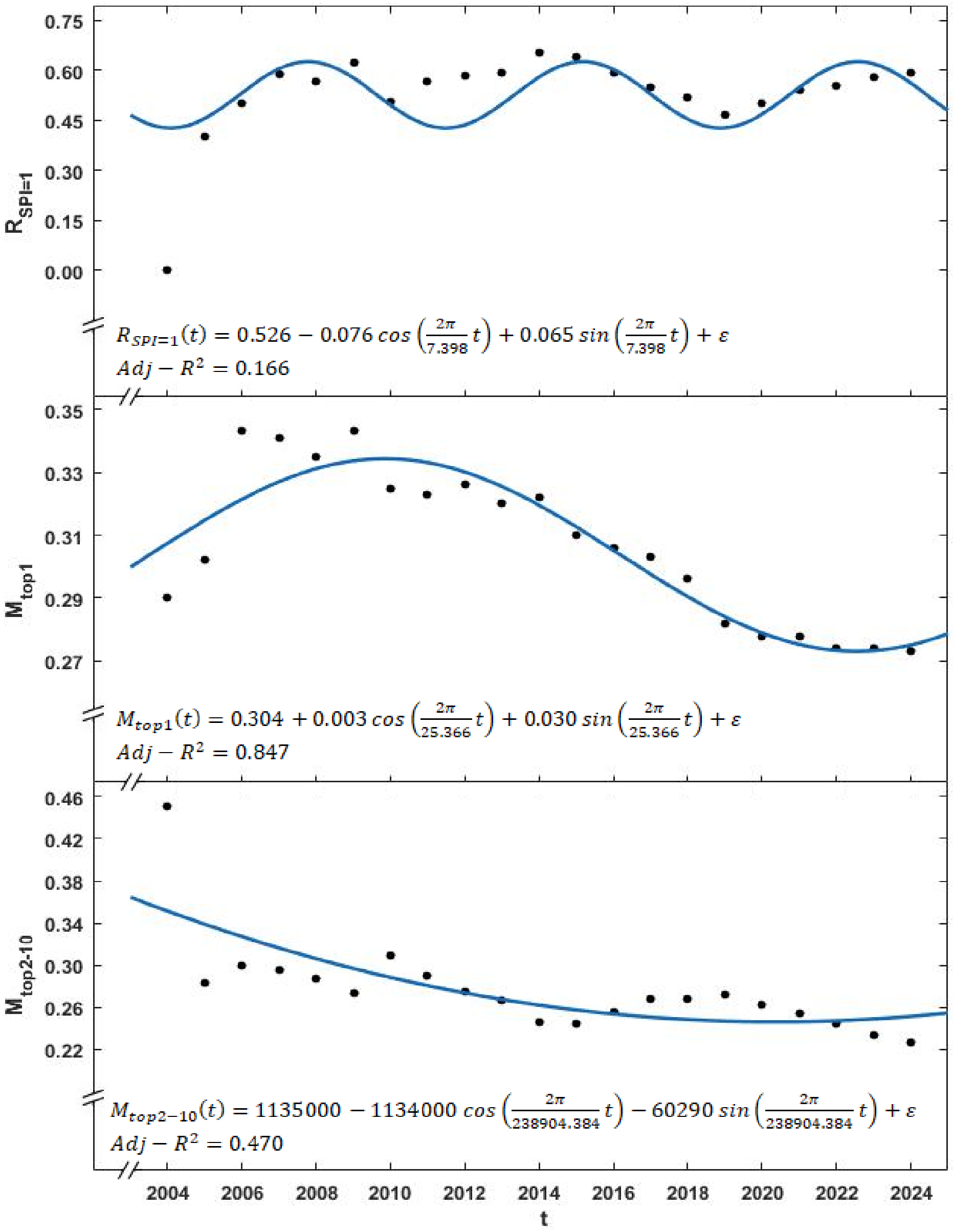

Figure A.3: Fitting of $R_{SPI=1}$, $M_{top1}$, and $M_{top2-10}$ on $t$ with a Fourier function in the first order for state-owned firms on the SME and GEM boards in 2004–2024.

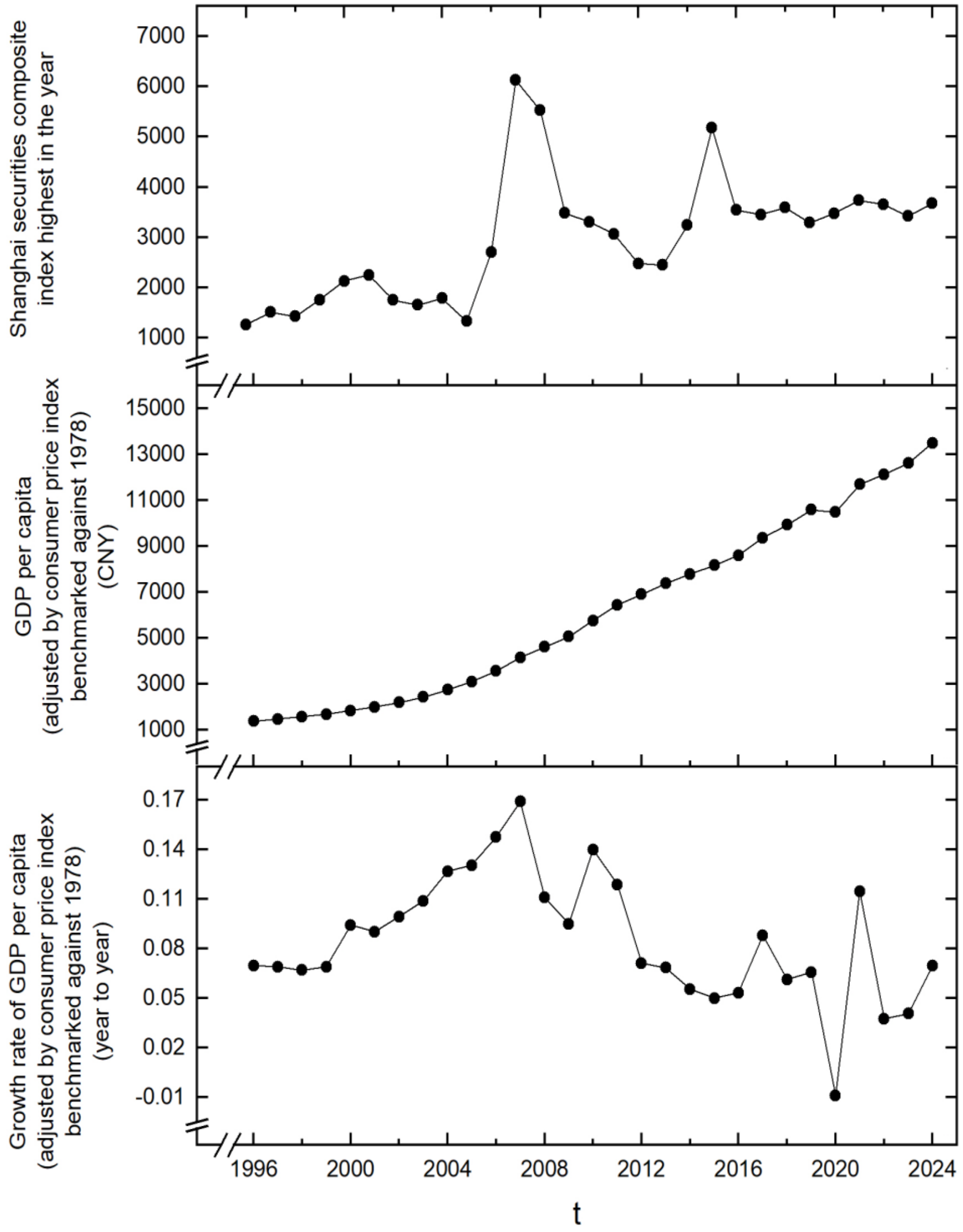

Figure A.4: Economic climate of China in 1996–2024.

## Appendix Tables:

Table A.1: Descriptions of the share rate of the top10 shareholders for privately owned firms on the main board.

| Year | Top1 Shareholder | | Top2-10 Shareholders | | $N_{sample}$ |
|---|---|---|---|---|---|
| | Mean | S.D. | Mean | S.D. | |
| 1992 | 0.382 | . | 0.268 | . | 1 |
| 1993 | 0.294 | 0.141 | 0.263 | 0.148 | 28 |
| 1994 | 0.300 | 0.118 | 0.262 | 0.141 | 32 |
| 1995 | 0.302 | 0.110 | 0.273 | 0.130 | 40 |
| 1996 | 0.316 | 0.114 | 0.250 | 0.130 | 87 |
| 1997 | 0.306 | 0.110 | 0.257 | 0.126 | 114 |
| 1998 | 0.298 | 0.105 | 0.275 | 0.122 | 124 |
| 1999 | 0.303 | 0.105 | 0.275 | 0.122 | 146 |
| 2000 | 0.297 | 0.105 | 0.276 | 0.118 | 166 |
| 2001 | 0.292 | 0.105 | 0.271 | 0.114 | 184 |
| 2002 | 0.287 | 0.100 | 0.280 | 0.118 | 226 |
| 2003 | 0.282 | 0.089 | 0.284 | 0.118 | 294 |
| 2004 | 0.284 | 0.089 | 0.286 | 0.118 | 334 |
| 2005 | 0.282 | 0.084 | 0.278 | 0.118 | 337 |
| 2006 | 0.273 | 0.089 | 0.246 | 0.118 | 389 |
| 2007 | 0.268 | 0.100 | 0.221 | 0.118 | 404 |
| 2008 | 0.265 | 0.105 | 0.207 | 0.118 | 395 |
| 2009 | 0.259 | 0.105 | 0.184 | 0.110 | 402 |
| 2010 | 0.254 | 0.100 | 0.186 | 0.114 | 412 |
| 2011 | 0.258 | 0.105 | 0.192 | 0.122 | 444 |
| 2012 | 0.258 | 0.105 | 0.196 | 0.126 | 452 |
| 2013 | 0.259 | 0.110 | 0.204 | 0.126 | 465 |
| 2014 | 0.264 | 0.110 | 0.214 | 0.126 | 504 |
| 2015 | 0.271 | 0.110 | 0.244 | 0.130 | 584 |
| 2016 | 0.276 | 0.114 | 0.270 | 0.130 | 660 |
| 2017 | 0.288 | 0.110 | 0.299 | 0.134 | 827 |
| 2018 | 0.288 | 0.110 | 0.297 | 0.130 | 873 |
| 2019 | 0.285 | 0.109 | 0.291 | 0.127 | 897 |
| 2020 | 0.280 | 0.106 | 0.297 | 0.128 | 963 |
| 2021 | 0.278 | 0.106 | 0.293 | 0.127 | 1029 |
| 2022 | 0.281 | 0.106 | 0.283 | 0.125 | 1038 |
| 2023 | 0.282 | 0.108 | 0.277 | 0.124 | 1045 |
| 2024 | 0.278 | 0.109 | 0.276 | 0.125 | 1068 |

Table A.2: The distribution of $\frac{top1\ holder's\ shares}{top2-10\ holder's\ shares}$ in the subsample $SPI = 1$ for privately owned firms on the main board.

| Year | $top1\ holder's\ shares/top2-10\ holder's\ shares$ | | | | | | | | | | | | | $N_{sample}$ |
|---|---|---|---|---|---|---|---|---|---|---|---|---|---|---|
| | [1.0,1.1) | [1.1,1.2) | [1.2,1.3) | [1.3,1.4) | [1.4,1.5) | [1.5,1.6) | [1.6,1.7) | [1.7,1.8) | [1.8,1.9) | [1.9,2.0) | …… | [10.5,10.6) | …… | |
| 1996 | 0.143 | 0.018 | 0.054 | 0.054 | 0.000 | 0.054 | 0.054 | 0.054 | 0.018 | 0.089 | …… | 0.000 | …… | 56 |
| 1997 | 0.094 | 0.047 | 0.031 | 0.078 | 0.063 | 0.063 | 0.047 | 0.016 | 0.031 | 0.031 | …… | 0.016 | …… | 64 |
| 1998 | 0.175 | 0.048 | 0.032 | 0.079 | 0.016 | 0.032 | 0.079 | 0.032 | 0.032 | 0.048 | …… | 0.000 | …… | 63 |
| 1999 | 0.111 | 0.097 | 0.028 | 0.028 | 0.083 | 0.056 | 0.083 | 0.014 | 0.042 | 0.028 | …… | 0.000 | …… | 72 |
| 2000 | 0.133 | 0.120 | 0.060 | 0.036 | 0.084 | 0.024 | 0.048 | 0.036 | 0.036 | 0.012 | …… | 0.000 | …… | 83 |
| 2001 | 0.156 | 0.067 | 0.111 | 0.022 | 0.111 | 0.044 | 0.011 | 0.033 | 0.078 | 0.000 | …… | 0.000 | …… | 90 |
| 2002 | 0.193 | 0.055 | 0.073 | 0.073 | 0.055 | 0.064 | 0.028 | 0.055 | 0.037 | 0.018 | …… | 0.000 | …… | 109 |
| 2003 | 0.146 | 0.108 | 0.069 | 0.069 | 0.069 | 0.023 | 0.046 | 0.046 | 0.023 | 0.008 | …… | 0.000 | …… | 130 |
| 2004 | 0.140 | 0.121 | 0.070 | 0.083 | 0.057 | 0.032 | 0.045 | 0.032 | 0.070 | 0.019 | …… | 0.000 | …… | 157 |
| 2005 | 0.154 | 0.083 | 0.077 | 0.058 | 0.090 | 0.038 | 0.045 | 0.032 | 0.064 | 0.032 | …… | 0.000 | …… | 156 |
| 2006 | 0.110 | 0.079 | 0.047 | 0.068 | 0.068 | 0.052 | 0.068 | 0.031 | 0.047 | 0.005 | …… | 0.000 | …… | 191 |
| 2007 | 0.091 | 0.083 | 0.061 | 0.065 | 0.052 | 0.043 | 0.039 | 0.030 | 0.030 | 0.009 | …… | 0.004 | …… | 230 |
| 2008 | 0.087 | 0.050 | 0.054 | 0.071 | 0.062 | 0.037 | 0.021 | 0.025 | 0.041 | 0.033 | …… | 0.004 | …… | 241 |
| 2009 | 0.082 | 0.060 | 0.060 | 0.064 | 0.026 | 0.034 | 0.030 | 0.026 | 0.037 | 0.045 | …… | 0.004 | …… | 267 |
| 2010 | 0.055 | 0.081 | 0.048 | 0.066 | 0.055 | 0.029 | 0.037 | 0.037 | 0.018 | 0.026 | …… | 0.000 | …… | 273 |
| 2011 | 0.082 | 0.082 | 0.061 | 0.044 | 0.024 | 0.044 | 0.031 | 0.027 | 0.034 | 0.027 | …… | 0.000 | …… | 294 |
| 2012 | 0.066 | 0.062 | 0.073 | 0.059 | 0.024 | 0.042 | 0.021 | 0.038 | 0.031 | 0.024 | …… | 0.000 | …… | 289 |
| 2013 | 0.061 | 0.075 | 0.082 | 0.044 | 0.041 | 0.058 | 0.024 | 0.038 | 0.020 | 0.027 | …… | 0.000 | …… | 293 |
| 2014 | 0.076 | 0.059 | 0.076 | 0.059 | 0.046 | 0.046 | 0.030 | 0.030 | 0.026 | 0.023 | …… | 0.000 | …… | 303 |
| 2015 | 0.099 | 0.081 | 0.081 | 0.068 | 0.047 | 0.050 | 0.043 | 0.037 | 0.056 | 0.025 | …… | 0.000 | …… | 322 |
| 2016 | 0.121 | 0.094 | 0.074 | 0.062 | 0.062 | 0.053 | 0.038 | 0.056 | 0.026 | 0.041 | …… | 0.000 | …… | 340 |
| 2017 | 0.107 | 0.094 | 0.104 | 0.064 | 0.056 | 0.069 | 0.074 | 0.064 | 0.033 | 0.041 | …… | 0.003 | …… | 393 |
| 2018 | 0.103 | 0.100 | 0.079 | 0.089 | 0.067 | 0.055 | 0.065 | 0.045 | 0.050 | 0.050 | …… | 0.000 | …… | 418 |
| 2019 | 0.113 | 0.094 | 0.090 | 0.071 | 0.055 | 0.062 | 0.057 | 0.044 | 0.034 | 0.032 | …… | 0.000 | …… | 435 |
| 2020 | 0.110 | 0.101 | 0.098 | 0.073 | 0.080 | 0.069 | 0.041 | 0.037 | 0.041 | 0.027 | …… | 0.000 | …… | 437 |
| 2021 | 0.106 | 0.090 | 0.088 | 0.088 | 0.077 | 0.061 | 0.040 | 0.035 | 0.040 | 0.029 | …… | 0.000 | …… | 479 |
| 2022 | 0.098 | 0.107 | 0.096 | 0.088 | 0.057 | 0.050 | 0.052 | 0.050 | 0.044 | 0.033 | …… | 0.000 | …… | 523 |
| 2023 | 0.095 | 0.083 | 0.098 | 0.064 | 0.079 | 0.057 | 0.042 | 0.051 | 0.040 | 0.042 | …… | 0.000 | …… | 529 |
| 2024 | 0.098 | 0.096 | 0.086 | 0.073 | 0.075 | 0.060 | 0.053 | 0.013 | 0.049 | 0.023 | …… | 0.000 | …… | 532 |

Table A.3: The distribution of $\frac{top1\ holder's\ shares}{top2-10\ holder's\ shares}$ in the subsample $SPI < 1$ for privately owned firms on the main board.

| Year | $top1\ holder's\ shares/top2-10\ holder's\ shares$ | | | | | | | | | | $N_{sample}$ |
|---|---|---|---|---|---|---|---|---|---|---|---|
| | [0.9,1.0) | [0.8,0.9) | [0.7,0.8) | [0.6,0.7) | [0.5,0.6) | [0.4,0.5) | [0.3,0.4) | [0.2,0.3) | [0.1,0.2) | [0,0.1) | |
| 1996 | 0.129 | 0.065 | 0.129 | 0.226 | 0.129 | 0.065 | 0.129 | 0.129 | 0.000 | 0.000 | 31 |
| 1997 | 0.140 | 0.140 | 0.160 | 0.240 | 0.060 | 0.140 | 0.060 | 0.060 | 0.000 | 0.000 | 50 |
| 1998 | 0.148 | 0.180 | 0.180 | 0.148 | 0.082 | 0.164 | 0.016 | 0.082 | 0.000 | 0.000 | 61 |
| 1999 | 0.162 | 0.189 | 0.135 | 0.189 | 0.095 | 0.135 | 0.068 | 0.027 | 0.000 | 0.000 | 74 |
| 2000 | 0.145 | 0.193 | 0.157 | 0.205 | 0.108 | 0.048 | 0.120 | 0.012 | 0.012 | 0.000 | 83 |
| 2001 | 0.160 | 0.170 | 0.160 | 0.181 | 0.138 | 0.117 | 0.032 | 0.032 | 0.011 | 0.000 | 94 |
| 2002 | 0.145 | 0.154 | 0.145 | 0.205 | 0.120 | 0.120 | 0.060 | 0.043 | 0.009 | 0.000 | 117 |
| 2003 | 0.165 | 0.152 | 0.152 | 0.238 | 0.079 | 0.116 | 0.061 | 0.024 | 0.012 | 0.000 | 164 |
| 2004 | 0.153 | 0.153 | 0.158 | 0.181 | 0.141 | 0.096 | 0.090 | 0.023 | 0.006 | 0.000 | 177 |
| 2005 | 0.177 | 0.133 | 0.199 | 0.177 | 0.133 | 0.105 | 0.061 | 0.011 | 0.006 | 0.000 | 181 |
| 2006 | 0.172 | 0.217 | 0.152 | 0.182 | 0.101 | 0.086 | 0.066 | 0.020 | 0.005 | 0.000 | 198 |
| 2007 | 0.132 | 0.155 | 0.218 | 0.126 | 0.161 | 0.086 | 0.057 | 0.034 | 0.029 | 0.000 | 174 |
| 2008 | 0.136 | 0.130 | 0.234 | 0.117 | 0.162 | 0.117 | 0.052 | 0.026 | 0.026 | 0.000 | 154 |
| 2009 | 0.207 | 0.126 | 0.163 | 0.126 | 0.163 | 0.104 | 0.052 | 0.037 | 0.022 | 0.000 | 135 |
| 2010 | 0.101 | 0.173 | 0.173 | 0.180 | 0.137 | 0.065 | 0.101 | 0.065 | 0.007 | 0.000 | 139 |
| 2011 | 0.100 | 0.180 | 0.173 | 0.113 | 0.187 | 0.107 | 0.067 | 0.067 | 0.007 | 0.000 | 150 |
| 2012 | 0.104 | 0.147 | 0.160 | 0.153 | 0.166 | 0.129 | 0.074 | 0.067 | 0.000 | 0.000 | 163 |
| 2013 | 0.087 | 0.157 | 0.163 | 0.099 | 0.215 | 0.105 | 0.087 | 0.081 | 0.006 | 0.000 | 172 |
| 2014 | 0.144 | 0.119 | 0.129 | 0.154 | 0.144 | 0.139 | 0.104 | 0.050 | 0.015 | 0.000 | 201 |
| 2015 | 0.130 | 0.137 | 0.134 | 0.153 | 0.141 | 0.153 | 0.099 | 0.038 | 0.015 | 0.000 | 262 |
| 2016 | 0.119 | 0.119 | 0.131 | 0.153 | 0.134 | 0.147 | 0.131 | 0.050 | 0.016 | 0.000 | 320 |
| 2017 | 0.154 | 0.101 | 0.131 | 0.157 | 0.143 | 0.134 | 0.101 | 0.067 | 0.012 | 0.000 | 434 |
| 2018 | 0.127 | 0.125 | 0.160 | 0.141 | 0.143 | 0.119 | 0.119 | 0.053 | 0.013 | 0.000 | 455 |
| 2019 | 0.106 | 0.134 | 0.171 | 0.128 | 0.134 | 0.143 | 0.117 | 0.058 | 0.009 | 0.000 | 462 |
| 2020 | 0.080 | 0.143 | 0.158 | 0.143 | 0.150 | 0.158 | 0.116 | 0.048 | 0.006 | 0.000 | 526 |
| 2021 | 0.096 | 0.140 | 0.140 | 0.155 | 0.142 | 0.162 | 0.100 | 0.056 | 0.009 | 0.000 | 550 |
| 2022 | 0.099 | 0.144 | 0.150 | 0.130 | 0.179 | 0.155 | 0.083 | 0.054 | 0.006 | 0.000 | 515 |
| 2023 | 0.118 | 0.136 | 0.128 | 0.145 | 0.174 | 0.163 | 0.079 | 0.052 | 0.004 | 0.000 | 516 |
| 2024 | 0.104 | 0.132 | 0.131 | 0.166 | 0.160 | 0.151 | 0.097 | 0.054 | 0.004 | 0.000 | 536 |

Table A.4: Descriptions of the share rate of the top10 shareholders for state-owned firms on the main board.

| Year | Top1 Shareholder | | Top2-10 Shareholders | | $N_{sample}$ |
|---|---|---|---|---|---|
| | Mean | S.D. | Mean | S.D. | |
| 1992 | 0.311 | 0.084 | 0.332 | 0.187 | 9 |
| 1993 | 0.288 | 0.148 | 0.222 | 0.141 | 70 |
| 1994 | 0.301 | 0.138 | 0.210 | 0.130 | 94 |
| 1995 | 0.318 | 0.114 | 0.240 | 0.134 | 117 |
| 1996 | 0.328 | 0.105 | 0.237 | 0.130 | 232 |
| 1997 | 0.329 | 0.105 | 0.224 | 0.126 | 314 |
| 1998 | 0.335 | 0.100 | 0.223 | 0.122 | 357 |
| 1999 | 0.332 | 0.100 | 0.234 | 0.126 | 391 |
| 2000 | 0.327 | 0.100 | 0.232 | 0.130 | 458 |
| 2001 | 0.327 | 0.100 | 0.227 | 0.130 | 502 |
| 2002 | 0.329 | 0.100 | 0.232 | 0.130 | 514 |
| 2003 | 0.333 | 0.100 | 0.234 | 0.130 | 508 |
| 2004 | 0.334 | 0.095 | 0.243 | 0.130 | 525 |
| 2005 | 0.328 | 0.095 | 0.241 | 0.130 | 540 |
| 2006 | 0.318 | 0.105 | 0.205 | 0.126 | 644 |
| 2007 | 0.316 | 0.110 | 0.193 | 0.126 | 661 |
| 2008 | 0.314 | 0.110 | 0.185 | 0.126 | 658 |
| 2009 | 0.313 | 0.105 | 0.186 | 0.126 | 648 |
| 2010 | 0.316 | 0.110 | 0.184 | 0.130 | 661 |
| 2011 | 0.315 | 0.105 | 0.184 | 0.130 | 651 |
| 2012 | 0.317 | 0.105 | 0.185 | 0.134 | 638 |
| 2013 | 0.317 | 0.105 | 0.191 | 0.134 | 625 |
| 2014 | 0.317 | 0.100 | 0.191 | 0.130 | 636 |
| 2015 | 0.319 | 0.100 | 0.206 | 0.130 | 651 |
| 2016 | 0.317 | 0.100 | 0.220 | 0.130 | 667 |
| 2017 | 0.318 | 0.100 | 0.225 | 0.130 | 681 |
| 2018 | 0.321 | 0.100 | 0.230 | 0.134 | 675 |
| 2019 | 0.320 | 0.099 | 0.235 | 0.136 | 683 |
| 2020 | 0.319 | 0.101 | 0.234 | 0.132 | 703 |
| 2021 | 0.321 | 0.102 | 0.229 | 0.129 | 730 |
| 2022 | 0.318 | 0.106 | 0.230 | 0.134 | 772 |
| 2023 | 0.315 | 0.105 | 0.231 | 0.135 | 788 |
| 2024 | 0.320 | 0.105 | 0.220 | 0.128 | 762 |

Table A.5: Descriptions of the share rate of the top10 shareholders for privately owned firms on the SME and GEM boards.

| Year | Top1 Shareholder | | Top2-10 Shareholders | | $N_{sample}$ |
|---|---|---|---|---|---|
| | Mean | S.D. | Mean | S.D. | |
| 2004 | 0.318 | 0.109 | 0.379 | 0.109 | 28 |
| 2005 | 0.294 | 0.093 | 0.309 | 0.104 | 37 |
| 2006 | 0.307 | 0.093 | 0.350 | 0.109 | 68 |
| 2007 | 0.316 | 0.095 | 0.359 | 0.106 | 127 |
| 2008 | 0.326 | 0.102 | 0.347 | 0.106 | 172 |
| 2009 | 0.320 | 0.102 | 0.335 | 0.105 | 238 |
| 2010 | 0.314 | 0.103 | 0.354 | 0.110 | 475 |
| 2011 | 0.313 | 0.102 | 0.350 | 0.107 | 670 |
| 2012 | 0.311 | 0.101 | 0.333 | 0.112 | 775 |
| 2013 | 0.305 | 0.102 | 0.304 | 0.112 | 785 |
| 2014 | 0.301 | 0.104 | 0.289 | 0.118 | 881 |
| 2015 | 0.292 | 0.104 | 0.293 | 0.119 | 1016 |
| 2016 | 0.283 | 0.104 | 0.296 | 0.116 | 1138 |
| 2017 | 0.283 | 0.103 | 0.309 | 0.119 | 1347 |
| 2018 | 0.278 | 0.102 | 0.302 | 0.115 | 1363 |
| 2019 | 0.271 | 0.103 | 0.291 | 0.115 | 1404 |
| 2020 | 0.266 | 0.104 | 0.278 | 0.115 | 1484 |
| 2021 | 0.262 | 0.105 | 0.272 | 0.116 | 1605 |
| 2022 | 0.260 | 0.106 | 0.264 | 0.118 | 1701 |
| 2023 | 0.259 | 0.107 | 0.255 | 0.117 | 1761 |
| 2024 | 0.258 | 0.109 | 0.245 | 0.119 | 1782 |

Table A.6: Descriptions of the share rate of the top10 shareholders for state-owned firms on the SME and GEM boards.

| Year | Top1 Shareholder | | Top2-10 Shareholders | | $N_{sample}$ |
|---|---|---|---|---|---|
| | Mean | S.D. | Mean | S.D. | |
| 2004 | 0.290 | 0.048 | 0.451 | 0.051 | 5 |
| 2005 | 0.302 | 0.101 | 0.283 | 0.155 | 10 |
| 2006 | 0.343 | 0.107 | 0.299 | 0.119 | 22 |
| 2007 | 0.341 | 0.109 | 0.295 | 0.123 | 39 |
| 2008 | 0.335 | 0.104 | 0.287 | 0.124 | 44 |
| 2009 | 0.343 | 0.100 | 0.274 | 0.124 | 53 |
| 2010 | 0.325 | 0.101 | 0.309 | 0.144 | 73 |
| 2011 | 0.323 | 0.096 | 0.290 | 0.129 | 85 |
| 2012 | 0.326 | 0.099 | 0.275 | 0.127 | 94 |
| 2013 | 0.320 | 0.096 | 0.267 | 0.128 | 96 |
| 2014 | 0.322 | 0.099 | 0.246 | 0.121 | 98 |
| 2015 | 0.310 | 0.099 | 0.245 | 0.115 | 106 |
| 2016 | 0.306 | 0.102 | 0.255 | 0.113 | 118 |
| 2017 | 0.303 | 0.097 | 0.268 | 0.117 | 126 |
| 2018 | 0.296 | 0.101 | 0.268 | 0.114 | 143 |
| 2019 | 0.282 | 0.097 | 0.272 | 0.111 | 185 |
| 2020 | 0.278 | 0.095 | 0.263 | 0.104 | 227 |
| 2021 | 0.278 | 0.094 | 0.254 | 0.112 | 269 |
| 2022 | 0.274 | 0.097 | 0.245 | 0.111 | 296 |
| 2023 | 0.274 | 0.098 | 0.234 | 0.111 | 305 |
| 2024 | 0.273 | 0.096 | 0.227 | 0.112 | 311 |

Table A.7: Correlation of $R_{SPI=1}$, $M_{top1}$, and $M_{top2-10}$ with the economic climate of China in 1996–2024 (Pearson correlation coefficient).

| | $R_{SPI=1}$ | $M_{top1}$ | $M_{top2-10}$ |
|---|---|---|---|
| Shanghai securities composite index highest in the year | 0.181<br>(0.347) | -0.535<br>(0.003) | -0.255<br>(0.181) |
| GDP per capita (adjusted by consumer price index benchmarked against 1978; CNY) | -0.118<br>(0.541) | -0.347<br>(0.065) | 0.170<br>(0.377) |
| Growth rate of GDP per capita (adjusted by consumer price index benchmarked against 1978; year to year) | 0.134<br>(0.487) | -0.248<br>(0.195) | -0.311<br>(0.100) |

*Notes*: Figures in the cells, from top to bottom, are the correlation coefficient and p-value.